\documentclass{appolb}
\usepackage{graphicx}
\usepackage{hyperref}
\usepackage{amsmath}
\usepackage{amssymb}
\usepackage{multirow}
\usepackage{float}
\usepackage{xcolor}   

\begin{document}

\title{Roper Resonance Structure and Exploration of Emergent Hadron Mass from CLAS Electroproduction Data
\thanks{Presented at ...}%
}
\author{Victor I. Mokeev and Daniel S. Carman
\address{Thomas Jefferson National Accelerator Facility, Newport News, Virginia 23606}}
\maketitle
\begin{abstract}
The $N(1440)1/2^+$ nucleon resonance, first identified in 1964 by L.D. Roper and collaborators in analyses of $\pi N$ hadroproduction data, has continued to provide pivotal insights that serve to advance our understanding of nucleon excited states. In this contribution, we present results from studies of the structure of the Roper resonance based on exclusive $\pi N$ and $\pi^+\pi^-p$ electroproduction data measured with the CLAS detector at Jefferson Lab. These analyses have revealed the Roper resonance as a complex interplay between an inner core of three dressed quarks and an external meson--baryon cloud. Analyses of the CLAS results on the evolution of the Roper resonance electroexcitation amplitudes with photon virtuality $Q^2$, within the framework of the Continuum Schwinger Method, have conclusively demonstrated the capability to gain insight into the strong interaction dynamics responsible for generating more than 98\% of hadron mass. Further extension of such studies to higher $Q^2$--through experiments currently underway with the CLAS12 detector and in the future with a potential CEBAF energy upgrade to 22 GeV--offers the only foreseeable opportunity to explore the full range of distances where the dominant portion of hadron mass and resonance structure emerge. 
\end{abstract}
  
\section{Introduction}

Since its discovery in the mid--1960s, the $N(1440)1/2^+$ (Roper) resonance, identified in analyses of $\pi N$ scattering data led by L.D. Roper~\cite{Roper:1964zza, Roper:1965pfb}, has served to provide essential insights into the dynamics of the strong interaction in the regime where the QCD running coupling constant $\alpha_s$ becomes comparable to unity. This so--called strong--QCD (sQCD) regime underlies the emergence of the nucleon resonance ($N^*$) spectrum and the structure of these excited states.

The Roper resonance exhibits several distinctive features:
\begin{itemize}
\item It is the first excited state of the nucleon with spin–parity $J^P = 1/2^+$. Its Breit–Wigner mass of 1.44 GeV is notably lower than that of the $N(1535)1/2^-$ chiral partner of the nucleon.
\item The $\approx$350~MeV decay width of the $N(1440)1/2^+$ is substantially larger than that of other $N^*$ states in the mass range up to 1.55~GeV, which are typically below 180~MeV.
\end{itemize}

The mass ordering between the $N(1440)1/2^+$ and $N(1535)1/2^-$ has long posed a challenge for constituent quark models (CQMs). These models incorporate interactions between quarks that are responsible for three–quark ($3q$) configuration mixing that typically originates from effective gluon exchange. In CQMs, the predicted mass of the chiral partner of the ground--state nucleon with $J^P=1/2^-$ is smaller than that of the first excited state with $J^P=1/2^+$, which is interpreted as the first radial excitation of the $3q$ system. This result holds across a variety of confining potentials~\cite{Giannini:2015zia}, as well as in relativistic~\cite{Capstick:1986ter, Capstick:1994ne, Capstick:2000qj} and effective field--theory quark models~\cite{Loring:2001kx}. This prediction, however, stands in stark contrast to the experimentally observed mass ordering between the $N(1440)1/2^+$ and $N(1535)1/2^-$~\cite{ParticleDataGroup:2024cfk}.

By introducing a more sophisticated interaction responsible for $3q$ configuration mixing--specifically, a combination of effective gluon--exchange and flavor--dependent terms--a significantly improved description of the $N^*$ spectrum has been achieved. This approach successfully reproduces the experimentally observed ordering of the $N(1440)1/2^+$ and $N(1535)1/2^-$ within both the CQM~\cite{Giannini:2015zia} and effective field–theory models~\cite{Ronniger:2011td}.

The octet of pseudoscalar mesons represents the Goldstone bosons associated with chiral symmetry breaking. The advancements in theory noted earlier highlight the critical importance of properly accounting for chiral symmetry breaking in describing the $N^*$ spectrum. The QCD--based framework for dynamical chiral symmetry breaking (DCSB) has been developed over the past decade within the Continuum Schwinger Method (CSM) approach~\cite{Achenbach:2025kfx, Burkert:2017djo, Ding:2022ows}.

Another peculiar feature of the $N(1440)1/2^+$ that has attracted attention for several decades is its large decay width. Early analyses of $\pi N$ scattering data~\cite{Arndt:1985vj} revealed an intricate structure of the partial waves in which the Roper resonance is formed, showing the presence of two poles in the complex energy plane. This observation has been confirmed in numerous subsequent studies~\cite{Arndt:2003if, ParticleDataGroup:2024cfk}, including global multi--channel analyses of $\pi N$ scattering and photoproduction data within coupled--channel approaches~\cite{Doring:2009yv, Julia-Diaz:2007qtz, Kamano:2010ud}. However, the splitting between the two poles is too small to account for its Breit--Wigner decay width. Such a large decay width strongly suggests a significant coupling to the final state meson--baryon channels and emphasizes the important role of coupled--channel effects in describing the Roper.

Detailed studies of the $N(1440)1/2^+$ have been carried out through global coupled--channel analyses of $\pi N$ scattering and photoproduction data by the Argonne–Osaka group~\cite{Julia-Diaz:2007qtz, Kamano:2010ud}. It was found that the two poles associated with the Roper originate from a {\it common} bare pole located on the real energy axis at a mass of 1.763~GeV. As the hadronic couplings of this bare state are increased toward the values determined from data fits, the single bare pole on the real axis evolves into three poles in the complex energy plane. Two of these poles move downward across the real energy axis, thereby generating the shift of the dressed Roper mass towards that observed in experiments.

The mass of the $N(1440)1/2^+$ has also been computed within the CSM framework as a bound state of three dressed quarks. This approach provides a direct connection to the QCD Lagrangian, since both the dressed--quark mass function (or propagator) and the $qq$ interaction amplitudes are deduced as solutions of the QCD equations of motion for the quark and gluon fields at distance scales where the transition from perturbative to strongly coupled QCD occurs~\cite{Burkert:2017djo, Segovia:2015hra}. The CSM prediction for the $N(1440)1/2^+$ mass, 1.73~GeV, is remarkably close to the bare mass deduced from the Argonne--Osaka coupled--channel analysis. Furthermore, the CSM results identify the $N(1440)1/2^+$ as the first radial excitation of dressed quarks, consistent with expectations from CQMs.

At present, the CSM accounts only for the contribution of the quark core to the resonance structure. Therefore, the close agreement between the CSM and coupled--channel analysis results strongly suggests that the structure of the $N(1440)1/2^+$ arises from the interplay between an inner core of three dressed quarks and an external meson--baryon cloud. The meson--baryon interactions within this cloud are responsible for shifting the bare core mass downward toward the experimentally observed value \cite{Burkert:2017djo}.

The lattice QCD (LQCD) results on the mass and structure of the Roper resonance remain a subject of debate. In LQCD evaluations~\cite{Liu:2016rwa, Liu:2014jua}, the $N(1440)1/2^+$ is treated as a $3q$ state that can couple to $5q$ configurations to account for meson--baryon dressing through channels such as $\pi N$, $\eta N$, and others. The computations of the nucleon and the $N(1440)1/2^+$ masses were performed for a pion mass of 330~MeV. After chiral extrapolation to the physical pion mass, this approach successfully reproduced the experimentally observed Roper mass and resolved the long--standing problem of the mass ordering between the $N(1440)1/2^+$ and $N(1535)1/2^-$. Owing to its chiral symmetry properties and its pattern for chiral symmetry breaking, this approach provides a natural explanation for the correct reproduction of the Roper mass~\cite{Burkert:2017djo, Liu:2016rwa}.

More recently, the nature of the Roper has been investigated by combining LQCD results with the Hamiltonian Effective Field Theory (HEFT) framework~\cite{Owa:2025mep}. In this approach, the $\pi N$ scattering amplitudes are derived from an effective meson--baryon Hamiltonian that incorporates the nucleon, the $N(1440)1/2^+$, and two--body meson–baryon channels such as $\pi N$, $\eta N$, $\pi\Delta$, and $\sigma N$. The HEFT amplitudes were simultaneously fit to the $\pi N$ scattering data and to the energy levels predicted by LQCD within a box of finite size. These studies concluded that the Roper can be understood predominantly as a dynamically generated state, with only a negligible contribution from the $3q$ core.

To distinguish between the different theoretical interpretations of the Roper resonance, studies of the $N^*$ masses alone are insufficient. As emphasized in the review of Ref.~\cite{Burkert:2017djo}, “Critical additional tests are imposed by requiring that the theoretical picture combine a prediction of the Roper’s mass with detailed descriptions of its structure and how that structure is revealed in the momentum dependence of the proton--Roper transition form factors.” Experimental measurements with the CLAS detector at Jefferson Lab (JLab) have provided results on the electroexcitation amplitudes of the Roper~\cite{Aznauryan:2009mx, Mokeev:2023zhq, Mokeev:2015lda, Mokeev:2012vsa}, enabling detailed exploration of its internal structure and the evolution of its active structural components across different distance scales.

\section{Roper Studies in Electroproduction Experiments with CLAS}
\label{elcpupl_p11}

Studies of exclusive meson electroproduction in the resonance region, carried out during the 6--GeV era with the CLAS detector, have provided the dominant part of the world’s data on most exclusive meson electroproduction channels~\cite{Mokeev:2022xfo}. For the first time, an extensive body of information--comprising $\approx$150,000 data points on differential cross sections and polarization asymmetries--has become available over the $W$ range from the pion threshold up to about 2~GeV for $Q^2$ from the photon point up to 5~GeV$^2$. The measured observables cover nearly the full angular range for final state hadron emission in the center--of--mass frame. The numerical results for all measured observables are stored in the CLAS Physics Database~\cite{Chesnokov:2022gjb, CLAS:DB}.

Analysis of these experimental results within reaction models made it possible to determine the $N^*$ electroexcitation amplitudes--referred to as the $\gamma_v p N^*$ electrocouplings, as defined in Ref.~\cite{Aznauryan:2011qj}--for most excited nucleon states with masses up to 1.75~GeV for $Q^2$ up to 5~GeV$^2$. The numerical results for these extracted electrocouplings are available online~\cite{gvnstarp:gwu, gvnstarp:dat}, as well as in Ref.~\cite{HillerBlin:2019jgp}. The first results on the $N^*$ electrocouplings from global multi--channel analyses of meson photo--, electro--, and hadroproduction by the groups from J{\"u}lich--Bonn--Washington \cite{Wang:2024byt} and Argonne--Osaka~\cite{Julia-Diaz:2007qtz,Kamano:2016bko} have become available. 

\subsection{Roper Structure from $\pi N$ and $\pi^+\pi^-p$ Electroproduction Data}
\label{elcoupl_p11exp}

\begin{figure*}[htbp]
\begin{center}
\includegraphics[width=1.0\textwidth]{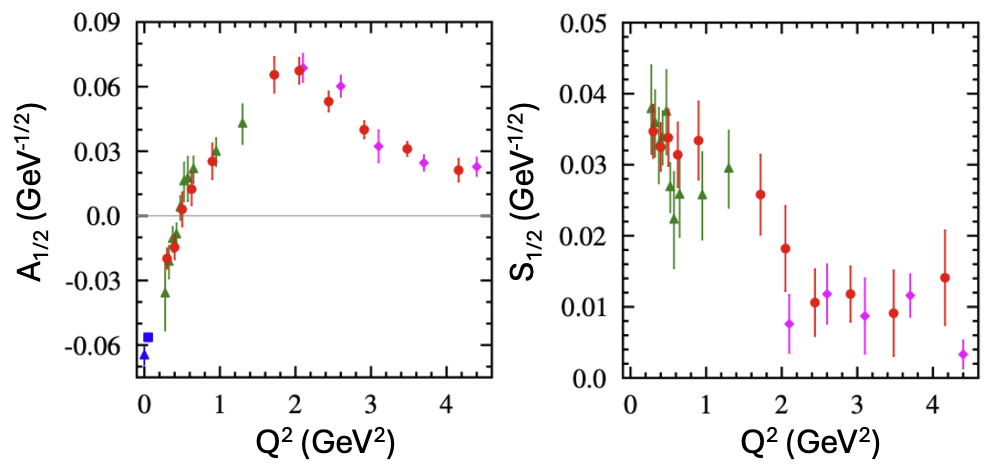}
\caption{The $\gamma_vpN^*$ electrocouplings for the $N(1440)1/2^+$ obtained from independent studies of $\pi N$ (red circles) \cite{Aznauryan:2011qj, Aznauryan:2009mx} and $\pi^+\pi^-p$ (green triangles and magenta diamonds) \cite{Mokeev:2015lda, Mokeev:2012vsa} electroproduction off protons. The photocouplings (in blue) are taken from Refs.~\cite{CLAS:2009tyz, ParticleDataGroup:2024cfk}.}
\label{roper_coupl_exp}
\end{center}
\end{figure*}

The $\gamma_v p N^*$ electrocouplings of the $N(1440)1/2^+$ have been determined from independent analyses of the two major meson electroproduction channels in the resonance region: $\pi N$~\cite{Aznauryan:2011qj, Aznauryan:2009mx} and $\pi^+\pi^-p$~\cite{Mokeev:2015lda, Mokeev:2012vsa}. These results for the $N(1440)1/2^+$ are shown in Fig.~\ref{roper_coupl_exp}.

\begin{figure*}[htbp]
\begin{center}
\includegraphics[width=0.6\textwidth]{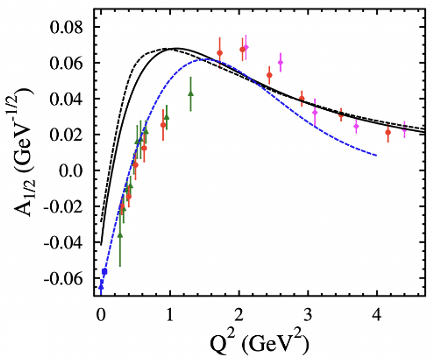}
\caption{Comparison of the $A_{1/2}$ electrocouplings of the $N(1440)1/2^+$ (see
Fig.~\ref{roper_coupl_exp} for details) to the computation from CSM \cite{Segovia:2015hra} (black solid line). The descriptions achieved within light--front quark models are also shown that i) implement a phenomenological momentum--dependent dressed quark mass~\cite{Aznauryan:2018okk} (black dashed line) and ii) account for both an inner core of three constituent quarks and an external meson--baryon cloud \cite{Obukhovsky:2011sc} (blue dashed line).} 
\label{roper_coupl_theor}
\end{center}
\end{figure*}

The electrocouplings of the Roper have been determined with good accuracy over the broad range of $Q^2$ from 0--5.0~GeV$^2$. The consistent results obtained from the independent analyses of the two major meson electroproduction channels with different non--resonant contributions provide strong evidence for the reliability of their extraction. A good description of the electrocouplings of the $N(1440)1/2^+$ has been achieved within the CSM framework~\cite{Segovia:2015hra} for $Q^2 > 2$~GeV$^2$. The CSM predictions are compared with the CLAS results for the $A_{1/2}$ electrocouplings in Fig.~\ref{roper_coupl_theor}.

\begin{figure}[htbp]
\begin{center}
\includegraphics[width=0.9\textwidth]{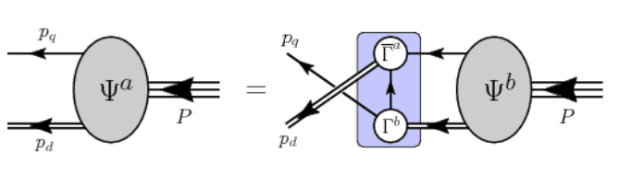}
\vspace{-1mm}
\caption{Faddeev equation for computation of the masses and wavefunctions of the quark core of the ground and excited states of the nucleon. The kernel for the matrix--valued integral equations is represented by the blue area.} 
\label{Faddeev_eq}
\end{center}
\end{figure}

The masses and wavefunctions of the nucleon ground state and the $N(1440)1/2^+$ are obtained as solutions of the Faddeev equations for a system of three dressed quarks in the quark–diquark approximation \cite{Barabanov:2020jvn,Segovia:2014aza, Segovia:2015hra}. The basic building blocks of the Faddeev equation—the dressed--quark propagators and the amplitudes of $qq$ correlations—are derived in connection with the QCD Lagrangian. In contrast to the static $qq$ correlations used in CQMs, the diquarks in the CSM framework are fully dynamical objects that continuously interact with the third quark, forming new correlated $qq$ pairs, as illustrated in Fig.~\ref{Faddeev_eq}.

The masses of the nucleon and its excited states with given spin–parity have been obtained as poles in the corresponding $J^P$ partial waves of the Faddeev amplitude for the three dressed quarks, from the solutions of the Faddeev equations shown in Fig.~\ref{Faddeev_eq}. The wavefunctions of the nucleon ground and excited states were extracted from the residues of the Faddeev amplitudes at the pole positions. A zero crossing observed in the Chebyshev moments of the components of the $N(1440)1/2^+$ Faddeev amplitudes demonstrates that the quark core of this resonance corresponds to a system of a dressed quark--bystander in its first radial excitation relative to the $qq$--correlated pair~\cite{Burkert:2017djo}.

The $\gamma_v p N^*$ electrocouplings are evaluated by considering the virtual--photon interaction with the electromagnetic currents of the dressed quark--diquark system, including transitions between diquarks of identical or different spin–parities. The calculations also account for the virtual--photon interaction at the vertex describing diquark decay and recombination into or from an uncorrelated quark pair, represented by the blue--shaded area in Fig.~\ref{Faddeev_eq}.

At present, the CSM accounts only for the contribution of the quark core to the $N^*$ structure. The successful description of the $N(1440)1/2^+$ electrocouplings achieved within the CSM for $Q^2 > 2$~GeV$^2$ suggests that, at the corresponding distance scales, the quark core provides the dominant contribution to the structure of the Roper.

Within the light--front quark model~\cite{Aznauryan:2012ec, Aznauryan:2018okk}, the $N(1440)1/2^+$ is treated as a bound system of three constituent quarks in their first radial excitation. The parameterized momentum dependence of the constituent quark mass is introduced to reproduce the experimental data on the nucleon elastic form factors. While this phenomenological momentum dependence cannot be connected to QCD, nevertheless using the same momentum--dependent quark mass, the model has achieved a reasonable description of the $\gamma_v p N^*$ electrocouplings for all $N^*$ states with masses up to 1.6~GeV.

As shown in Fig.~\ref{roper_coupl_theor}, both the CSM and the light--front quark model provide a good description of the $N(1440)1/2^+$ electrocouplings for $Q^2$ from 2--5~GeV$^2$. This agreement demonstrates that for $Q^2 > 2$~GeV$^2$, the quark core is the dominant contributor to the structure of the $N(1440)1/2^+$. However, both the CSM~\cite{Segovia:2015hra} and the light--front quark model of Ref.~\cite{Aznauryan:2018okk} fail to reproduce the experimental results for the $N(1440)1/2^+$ electrocouplings at $Q^2 < 1$~GeV$^2$. This discrepancy indicates the presence of additional contributions to the resonance structure that become important at distance scales comparable to the baryon size. These contributions originate from the meson--baryon cloud.

The light--front quark model of Ref.~\cite{Obukhovsky:2011sc}, which incorporates both the quark--core and meson–baryon cloud components in the $N(1440)1/2^+$ structure, provides a much improved description of the data at $Q^2 < 1$~GeV$^2$, while maintaining a reasonable agreement at higher $Q^2$. This result explains the success of models that include only meson--baryon degrees of freedom in describing the $N(1440)1/2^+$ electrocouplings at low photon virtualities~\cite{Bauer:2014cqa, Krehl:1999km, Speth:2000zf}.

Analysis of the CLAS results on the $Q^2$ evolution of the Roper electrocouplings has revealed that its structure arises from a complex interplay between an inner core of three dressed quarks in the first radial excitation and an external meson--baryon cloud. At distance scales comparable to the hadron size, corresponding to $Q^2 < 1$~GeV$^2$, the meson–baryon cloud plays a significant role in the resonance structure. At higher photon virtualities, $Q^2 > 2$~GeV$^2$, the virtual photons penetrate the meson--baryon cloud and interact predominantly with the inner core of three dressed quarks.

\subsection{Further Evidence for Quark Core Contributions to Roper Structure}
\label{hadr_dec_p11}

In the extraction of the resonance parameters from the $\pi^+\pi^-p$ electroproduction data~\cite{Mokeev:2023zhq, Mokeev:2015lda, Mokeev:2012vsa}, we simultaneously varied the $\gamma_v p N^*$ electrocouplings of the nucleon resonances, their masses, total decay widths ($\Gamma_{tot}$), and partial decay widths to the $\pi\Delta$ and $\rho p$ final states. These parameters were fit to nine one--fold differential $\pi^+\pi^-p$ electroproduction cross sections in each ($W$, $Q^2$) bin covered in the analyses, using the JLab--Moscow State University JM23 reaction model~\cite{Mokeev:2023zhq, Mokeev:2008iw, Mokeev:2012vsa, Ripani:2000rz}.

The resonance parameters determined from the data fits include the electrocouplings, the partial decay widths into $\pi\Delta$ and $\rho p$, and the total decay widths. For each ($W$, $Q^2$) bin, these quantities were obtained as averages from the set of fits selected according to specific limits on the reduced $\chi^2$. These criteria were chosen to ensure that the fitted $\pi^+\pi^-p$ differential cross sections remained within the experimental uncertainties for most of the measured data points.

The total experimental uncertainties were evaluated as the quadratic sum of the statistical and kinematic--dependent systematic uncertainties. The mean values for the resonance parameters selected in the data fits were treated as the experimentally extracted values. The root--mean--square dispersions of the resonance parameters across the selected fits were taken as their respective uncertainties. The $N(1440)1/2^+$ mass, along with its total and partial decay widths to the $\pi\Delta$ and $\rho p$ final states, as determined from the fits to the $\pi^+\pi^-p$ electroproduction data, are listed in Table~\ref{hadr_n1440}.

\begin{table*}
\begin{center}
\begin{tabular}{|c|c|c|c|c|c|c|} \hline
$Q^2$ Interval & Mass            & $\Gamma_{tot}$ & $\Gamma_{\pi\Delta}$ & BF$_{\pi\Delta}$ & $\Gamma_{\rho p}$ & BF$_{\rho p}$  \\
(GeV$^2)$      & (GeV)           & (MeV)          & (MeV)                & \%               & (MeV)             & \% \\ \hline
0.25--0.60      & 1.458$\pm$0.012 & 363$\pm$39     & 142$\pm$48           & 23-58            & 6$\pm$4           & $<$2 \\ \hline
0.5--1.5        & 1.450$\pm$0.011 & 352$\pm$37     & 120$\pm$41           & 20-52            & 5$\pm$2           & $<$2 \\ \hline
2.0--3.5        & 1.457$\pm$0.008 & 331$\pm$54     & 129$\pm$52           & 20-65            & 6$\pm$2           & 1.1--2.6 \\ \hline
3.0--5.0        & 1.446$\pm$0.013 & 352$\pm$33     & 151$\pm$32           & 31-57            & 5$\pm$1           & 1.2--2.0 \\ \hline
\end{tabular}
\caption{Masses and total/partial hadronic decay widths of the $N(1440)1/2^+$ into the $\pi\Delta$ and $\rho p$ final states, determined from fits to the $\pi^+\pi^-p$ electroproduction cross sections performed independently in different $Q^2$ intervals. The results from Ref.~\cite{Mokeev:2023zhq} are shown in the last two rows, while those in the upper rows are taken from previous studies~\cite{Mokeev:2015lda, Mokeev:2012vsa} of $\pi^+\pi^-p$ electroproduction.}
\label{hadr_n1440} 
\end{center}
\end{table*}

The extracted parameters have become available from independent data fits within the $Q^2$ intervals listed in the left column of Table~\ref{hadr_n1440}. Over the broad range of $Q^2$ up to 5~GeV$^2$, a successful description of the data has been achieved using $Q^2$--independent masses, and total and partial decay widths of the $N(1440)1/2^+$. This result suggests that the $N(1440)1/2^+$ is excited in the $s$--channel through the $\gamma_v p$ interaction, as for $s$--channel resonances, the electroexcitation and hadronic decay amplitudes into different final states are independent.

Analyses of the CLAS meson electroproduction data have shown that resonances excited in the $s$--channel possess a quark core. Therefore, the observed $Q^2$--independent mass, total, and partial decay widths of the Roper resonance, as deduced from the $\pi^+\pi^-p$ electroproduction data, provides additional evidence that the structure of this state originates from the interplay between an inner core of three dressed quarks and an external meson--baryon cloud.

The results from studies of the $N(1440)1/2^+$ structure based on analyses of meson electroproduction data with CLAS support the predictions of the CSM~\cite{Segovia:2015hra} and the CQMs~\cite{Aznauryan:2018okk, Obukhovsky:2011sc}, while showing tension with results obtained from the combined HEFT and LQCD approach~\cite{Owa:2025mep}, which suggests that the Roper is dynamically generated through processes other than $s$--channel resonance excitation. However, the analysis of Ref.~\cite{Owa:2025mep} imposes a constraint that the mass of the three--quark state with $J^P = 1/2^+$ lies above 2~GeV.

It therefore cannot be excluded that a molecular--type meson--baryon state may exist in place of the $N(1440)1/2^+$, as proposed in Ref.~\cite{Wang:2023snv}, or that this state contains a significant $5q$ configuration component~\cite{Zou:2003zn}. In either case, this state would still be excited in the $s$--channel of the $\gamma_v p$ interaction, and its hadronic decay widths would remain $Q^2$--independent.

To discriminate among these alternative interpretations of the Roper, it is essential to obtain theoretical predictions for the $Q^2$ evolution of its electrocouplings and to confront them directly with the results extracted from meson electroproduction data analyses.

\section{Roper Electroexcitation and Emergence of Hadron Mass}
\label{p11_EHM}

The emergence of hadron mass (EHM) remains one of the most profound and unresolved problems within the Standard Model (SM). The issue becomes evident when comparing the measured masses of the proton and neutron with the sum of the masses of their partonic quark constituents. In the SM, the renormalization group–invariant (RGI) current masses of quarks are generated through the Higgs mechanism~\cite{Englert:2014zpa, Higgs:2014aqa}, a process confirmed by the discovery of the Higgs boson at CERN~\cite{ATLAS:2012yve, CMS:2012qbp}. This mechanism is responsible for generating the Lagrangian masses of the most fundamental constituents of matter known to date--quarks and leptons.

However, the Higgs mechanism contributes negligibly to the generation of nucleon and nucleon--excited--state ($N^*$) masses. Protons and neutrons are bound states composed of three light quarks, $uud$ and $udd$, respectively. The sum of the Higgs--generated current masses of these quarks accounts for less than 2\% of the observed nucleon mass. The comparison summarized in Table~\ref{nucleon_mass} employs the quark current masses from the Particle Data Group~\cite{ParticleDataGroup:2024cfk} evaluated at the renormalization scale $\zeta = 2.0$ GeV. This comparison makes it evident that the dominant share of the nucleon mass arises from mechanisms other than those associated with the Higgs field.

\begin{table*}
\setlength{\tabcolsep}{6pt} 
\renewcommand{\arraystretch}{1.3} 
\begin{tabular}{|c|c|c|} \hline
                                      & Proton         & Neutron \\ \hline
\multirow{2}{*}{Measured mass (MeV)}  & 938.2720813             & 939.5654133 \\
                                      & $\pm$ 0.0000058         & $\pm$ 0.0000058 \\  \hline      
Sum of the current quark masses (MeV) & 8.09$^{+1.45}_{-0.65}$  & 11.50$^{+1.45}_{-0.60}$ \\ \hline
Contribution of the current           & \multirow{2}{*}{$<$1.1} & \multirow{2}{*}{$<$1.4}\\ 
quark masses to $M_N$ (\%)            &                         &    \\ \hline
\end{tabular}
\caption{Comparison between the measured masses of the proton and neutron and the sum of the current quark masses of their three $u$ and $d$ valence quark constituents \cite{ParticleDataGroup:2024cfk}. Current quark masses are listed at a scale of 2\,GeV, but the comparison remains practically unchanged if renormalization group--invariant (RGI) current masses are used.}
\label{nucleon_mass}
\end{table*}

Substantial progress has been achieved over the past decade toward understanding EHM within the framework of CSM. The concept that more than 98\% of hadron mass arises dynamically through EHM is strongly supported by experimental studies of hadron structure in both the meson and baryon sectors, including the ground and excited states of the nucleon~\cite{Achenbach:2025kfx, Binosi:2025kpz, Ding:2022ows, Horn:2016rip}. At present, the CSM provides the only QCD--connected approach capable of predicting the $Q^2$ evolution of nucleon resonance electroexcitation amplitudes within the same theoretical framework that successfully describes the properties of a broad range of hadrons, including Nature’s most fundamental Nambu--Goldstone bosons. The development of other nonperturbative theoretical approaches to QCD capable of delivering a comparable array of results is still eagerly awaited.

Within the CSM, it has been demonstrated that the dominant part of the masses of both the ground and excited states of the nucleon is generated by strong QCD interactions in the transition region between perturbative QCD (pQCD) and sQCD. In this domain, the QCD running coupling $\alpha_s/\pi$ evolves from small values characteristic of the pQCD regime to values of order unity in the sQCD regime. At the distance scales corresponding to this transition, the effective degrees of freedom--dressed quarks and dressed gluons--emerge as the building blocks of hadronic matter.

\begin{figure}[htpb]
\begin{center}
\includegraphics[width=0.97\textwidth]{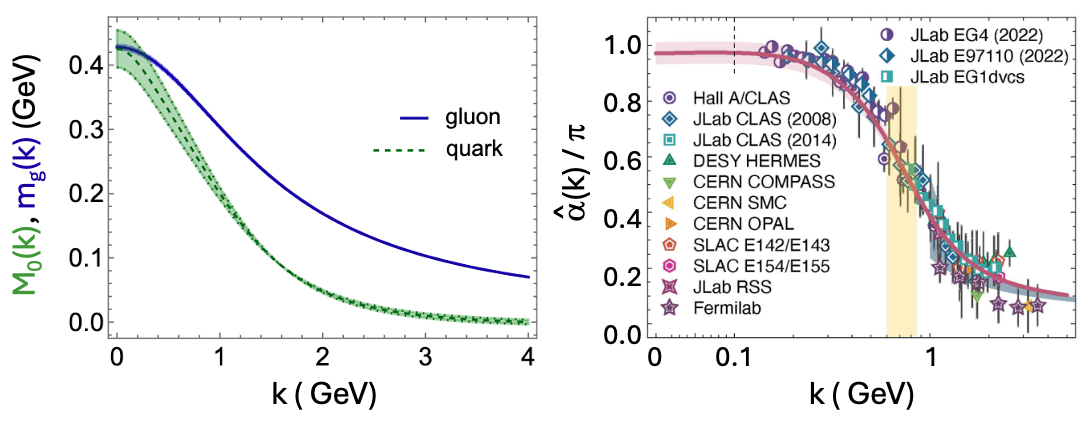}
\end{center}
\caption{(Left) CSM predictions for the momentum ($k=\sqrt{k^2}$) dependence of the dressed gluon (solid blue curve) and quark (dot--dashed green) mass functions in the chiral limit~\cite{Roberts:2020hiw,Roberts:2021xnz, Roberts:2021nhw}. For the quark, the associated band expresses existing uncertainties in the CSM predictions. (Right) CSM prediction~\cite{Cui:2019dwv} for the momentum dependence of the process--independent QCD running coupling, $\hat\alpha(k)$ (purple curve and associated uncertainty band, which includes uncertainties associated with the gluon mass function) compared with empirical results~\cite{Deur:2022msf} for the process--dependent effective charge defined via the Bjorken sum rule. The vertical yellow band marks the window of sQCD $\leftrightarrow$ pQCD transition in the running coupling. (A complete discussion of effective charges is available elsewhere~\cite{Deur:2023dzc}.  All sources of the data in the right panel are listed in Refs.~\cite{Deur:2023dzc, Deur:2022msf}.)}
\label{Quark_Gluon_massfunct}
\end{figure}

Dressed gluons, owing to the self--interaction terms encoded in the QCD Lagrangian, acquire a momentum--(or distance--)dependent emergent mass. The running gluon mass arises from initially massless QCD gluons through the transverse part of the dressed--gluon polarization tensor via the so--called Schwinger mechanism~\cite{Ding:2022ows, Schwinger:1962tn, Schwinger:1962tp}, which preserves gauge invariance. The momentum dependence of the dressed--gluon mass, computed from the QCD Lagrangian as a solution of the QCD equations of motion for the quark and gluon fields, is shown in Fig.~\ref{Quark_Gluon_massfunct} (left). The blue--shaded band in the figure indicates the uncertainties associated with the CSM evaluation.

There are essential differences in the evolution of the running dressed--gluon mass across the pQCD ($k > 2.0$~GeV), sQCD ($k < 0.5$~GeV), and transition ($0.5~{\rm GeV} < k < 2.0~{\rm GeV}$) domains. In this transition region, the running gluon mass increases rapidly as $k$ decreases, reaching values of approximately 0.4~GeV and thereby establishing the characteristic hadron mass scale. In contrast, within both the pQCD and sQCD regimes, the running gluon mass is nearly independent of $k$.

The momentum--dependent dressed--gluon mass determines the behavior of the QCD running coupling $\hat\alpha(k)$, shown in Fig.~\ref{Quark_Gluon_massfunct} (right). This coupling was evaluated using a momentum--subtraction renormalization scheme, such that the RGI scale is $\Lambda_{\rm QCD} = 0.52$~GeV for $n_f = 4$. A remarkable feature is observed in Fig.~\ref{Quark_Gluon_massfunct} (right): while the pQCD running coupling exhibits a Landau pole--diverging at $\Lambda_{\rm QCD}$, regardless of the loop order in the calculation (see discussion in Ref.~\cite{Deur:2023dzc})—$\hat\alpha(k)$, computed within CSM is a smooth, finite function for all $k^2 \ge 0$. The emergence of a dynamical gluon mass scale removes the Landau pole, highlighting the central role of EHM in generating a well--defined, infrared--finite QCD coupling~\cite{Achenbach:2025kfx, Ding:2022ows}.

When the QCD running coupling $\alpha_s/\pi$ becomes comparable to unity, the energy stored in the dressed gluon fields is converted into the momentum--dependent mass of dressed quarks. The corresponding dressed--quark mass function, computed from QCD within the CSM framework~\cite{Ding:2022ows, Roberts:2020hiw}, is shown in Fig.~\ref{Quark_Gluon_massfunct} (left). The green--shaded band represents the uncertainty associated with the CSM evaluation. The mass of a bare quark increases rapidly with distance in the pQCD $\leftrightarrow$ sQCD transition region, eventually reaching values comparable to the hadronic mass scale defined by the running gluon mass in the infrared.

The strong interaction among three dressed quarks gives rise to both the ground state and the spectrum of excited nucleon states within the mass range from 1~GeV to 2.5~GeV. This range is consistent with the hadron mass scale of fully dressed quarks of around 0.4 GeV (see Fig.~\ref{Quark_Gluon_massfunct}). The three pillars of EHM revealed within the CSM framework--(a) the running dressed--gluon mass, (b) the QCD running coupling $\alpha_s$, and (c) the running mass of dressed quarks--collectively account for the generation of more than 98\% of the hadron mass. The emergent hadron mass is produced entirely through strong interaction dynamics encoded in the QCD Lagrangian, which governs the transformation of the fundamental QCD degrees of freedom into the effective degrees of freedom in the transition towards the sQCD regime, namely, dressed quarks and gluons. Consequently, there is no need for any additional fields, such as the Higgs field, to explain the origin of the dominant portion of hadron mass.

\begin{figure}[t]
\begin{center}    
\includegraphics[width=1.0\textwidth]{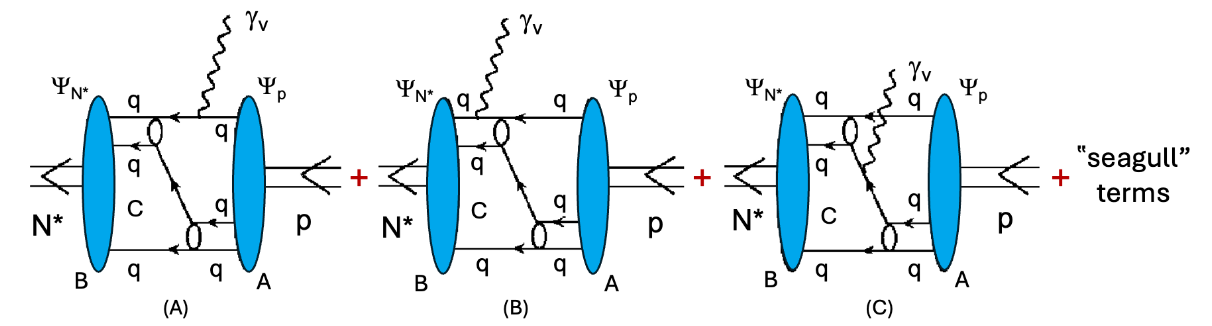}
\end{center}
\caption{CSM description of resonance electroexcitation amplitudes~\cite{Barabanov:2020jvn, Burkert:2017djo, Cloet:2013jya, Segovia:2014aza, Segovia:2015hra}. All diagrams describe the transition $p \to$ dressed quark plus interacting diquark correlations $\to$ $N^*$. The Faddeev amplitudes for the transitions between dressed quark + diquark configurations and the ground or excited states of the nucleon correspond to the proton, $\psi_p$, or $N^*$, $\psi_{N^*}$, wavefunctions, respectively.}
\label{diagdse}
\end{figure}

Within the framework of CSMs, nucleon resonance electroexcitation is described by an electromagnetic--current--conserving set of diagrams~\cite{Barabanov:2020jvn, Burkert:2017djo, Cloet:2013jya, Segovia:2014aza, Segovia:2015hra}, partially illustrated in Fig.~\ref{diagdse}. These diagrams represent the dominant contributions to the resonance electroexcitation amplitudes, while the complete gauge--invariant set of transition amplitudes can be found in Ref.~\cite{Segovia:2014aza}, Fig.~C1. The amplitudes that describe the interaction of real or virtual photons with dressed quarks and diquark correlations are directly sensitive to the dressed--quark propagators. The mass function of the dressed quark constitutes a key element of the propagator. Consequently, studying the $Q^2$ evolution of the $\gamma_v p N^*$ electrocouplings provides a powerful means to map the momentum dependence of the dressed--quark mass. Achieving a successful description of the electroexcitation amplitudes of various nucleon excited states using a common dressed--quark mass function--computed within the CSM directly from the QCD Lagrangian--would offer nearly model--independent evidence for insight into the momentum dependence of the dressed--quark mass.

\begin{figure}[h]
\centering    
\includegraphics[width=0.92\textwidth]{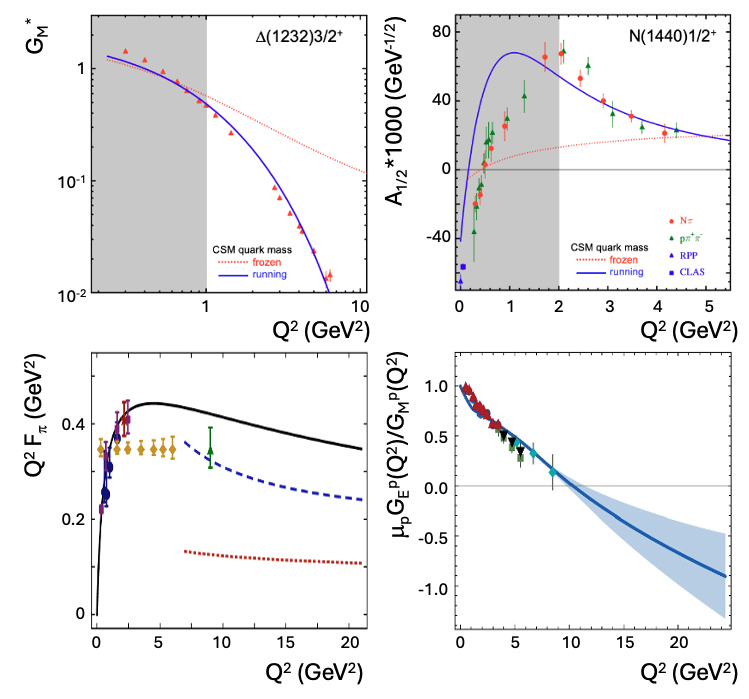}
\caption{(Top left) Description of the $N \to \Delta$ magnetic transition form factor normalized to the dipole fit $G^*_M/3G_D$ with $G_D(Q^2)=(1+Q^2/0.71\,{\rm GeV}^2)^{-2}$ and (top right) of the $N(1440)1/2^+$ $A_{1/2}$ electrocoupling achieved using CSMs~\cite{Segovia:2014aza, Segovia:2015hra, Wilson:2011aa}. Results obtained with a momentum--independent (frozen) dressed quark mass~\cite{Segovia:2013uga, Wilson:2011aa} (dotted red curves) are compared with QCD--kindred results (solid blue curves) obtained with a momentum--dependent quark mass function of the type drawn in Fig.~\ref{Quark_Gluon_massfunct}~\cite{Segovia:2014aza, Segovia:2015hra}. The data were taken from Refs.~\cite{CLAS:2006fml, Aznauryan:2009mx, Villano:2009sn} for $\pi N$ and Refs.~\cite{Mokeev:2023zhq, Mokeev:2015lda, CLAS:2012wxw} for $\pi^+\pi^-p$. The photocouplings for the $N(1440)1/2^+$ are from the PDG \cite{ParticleDataGroup:2024cfk} and Ref.~\cite{CLAS:2009tyz}, blue square and triangle, respectively. The approximate range of $Q^2$ where the contributions from the meson--baryon cloud remain substantial are highlighted in gray. (Bottom left) A successful description of the pion elastic form factor~\cite{NA7:1986vav, Horn:2007ug, JeffersonLab:2008jve} and (bottom right) the ratio of the proton electric--to--magnetic form factors, $\mu_p G_E(Q^2)/G_M(Q^2)$~\cite{JeffersonLabHallA:2001qqe, JeffersonLab:2008jve, JeffersonLabHallA:1999epl, Punjabi:2005wq} has been achieved within the CSM framework~\cite{Cui:2020rmu} using the same dressed--quark mass function that successfully describes both the pion elastic and the $N \to N^*$ transition form factors/electrocouplings.}
\label{csm_delta_roper}
\end{figure}

To elucidate the impact of the running quark mass, the CSM predictions for the Roper resonance electrocouplings were obtained using two different forms of the $qq$ interaction: ({a}) a simplified, momentum--independent $qq$ contact interaction~\cite{Segovia:2013uga, Wilson:2011aa}, and ({b}) Schwinger functions that express a momentum dependence consistent with a $qq$ interaction derived from the QCD Lagrangian~\cite{Segovia:2014aza, Segovia:2015hra}. In the simplified case ({a}), the interaction kernel for the gap equation is treated within the rainbow--ladder (RL) truncation, which represents the leading--order term in a symmetry--preserving and systematically improvable approximation scheme~\cite{Bender:1996bb, Munczek:1994zz}. Using this simplified $qq$ interaction--also referred to as the symmetry--preserving contact--interaction (SCI) approximation--yields a momentum--independent dressed--quark mass $M \approx 0.35$ GeV, consistent with the infrared behavior of the running quark mass displayed in Fig.~\ref{Quark_Gluon_massfunct} (left). The SCI predictions for the $A_{1/2}$ electrocoupling of the $N(1440)1/2^+$ are shown as red dashed curves in Fig.~\ref{csm_delta_roper}~\cite{Segovia:2013uga, Wilson:2011aa}. As seen in Fig.~\ref{csm_delta_roper} (top right), apart from reproducing the zero crossing in the correct region, the SCI fails to describe the $N(1440)1/2^+$ electrocouplings across the entire range of $Q^2$ covered by the measurements.

In contrast, approach ({b})--built upon the dressed quark mass function that encodes the momentum dependence generated by QCD’s $qq$ interaction~\cite{Segovia:2014aza, Segovia:2015hra}—yields the solid blue curves in Fig.~\ref{csm_delta_roper} (top row) for both the Roper and $\Delta(1232)3/2^+$. In this case, the electroexcitation amplitudes of both resonances are well described over the entire range of $Q^2$ where the quark--core contributions are expected to dominate. The experiment--theory comparisons for the $N^*$ electroexcitation amplitudes in Fig.~\ref{csm_delta_roper} (top row) provide compelling evidence that {\it the mass of dressed quarks runs with distance}. Additional support for the role of constituent quarks with running masses as the active structural degrees of freedom in $N^*$ states has been obtained from quark--model analyses of $\gamma_v pN^*$ electrocouplings~\cite{Aznauryan:2012ec, Aznauryan:2016wwm, Aznauryan:2018okk}. Significantly, within CSM the successful simultaneous description of the JLab data on the $\Delta(1232)3/2^+$ and $N(1440)1/2^+$ electroexcitation amplitudes was achieved using the same dressed--quark mass function that also reproduces the pion and nucleon elastic electromagnetic form factors~\cite{Yao:2024uej, Yao:2024drm}, as shown in Fig.~\ref{csm_delta_roper} (bottom row).

Consistent results for the dressed--quark mass function, obtained from independent studies of the pion and ground--state nucleon structure, as well as from the $\gamma_v p N^*$ electroexcitation amplitudes of states with different structures—namely, the spin--isospin flip in the $\Delta(1232)3/2^+$ and the first radial excitation of dressed quarks in the $N(1440)1/2^+$--provide compelling evidence for the existence of dynamically generated dressed quarks with running masses. These quarks emerge as the active constituents in the structure of pions, kaons, ground--state nucleons, and the $\Delta(1232)3/2^+$ and $N(1440)1/2^+$. This success further demonstrates the capability of mapping the RGI dressed--quark mass function directly from the experimental data on the $Q^2$ evolution of the $N^*$ electroexcitation amplitudes when analyzed within the framework of CSMs.

\begin{figure}[htbp]
\centering
\includegraphics[width=0.85\columnwidth]{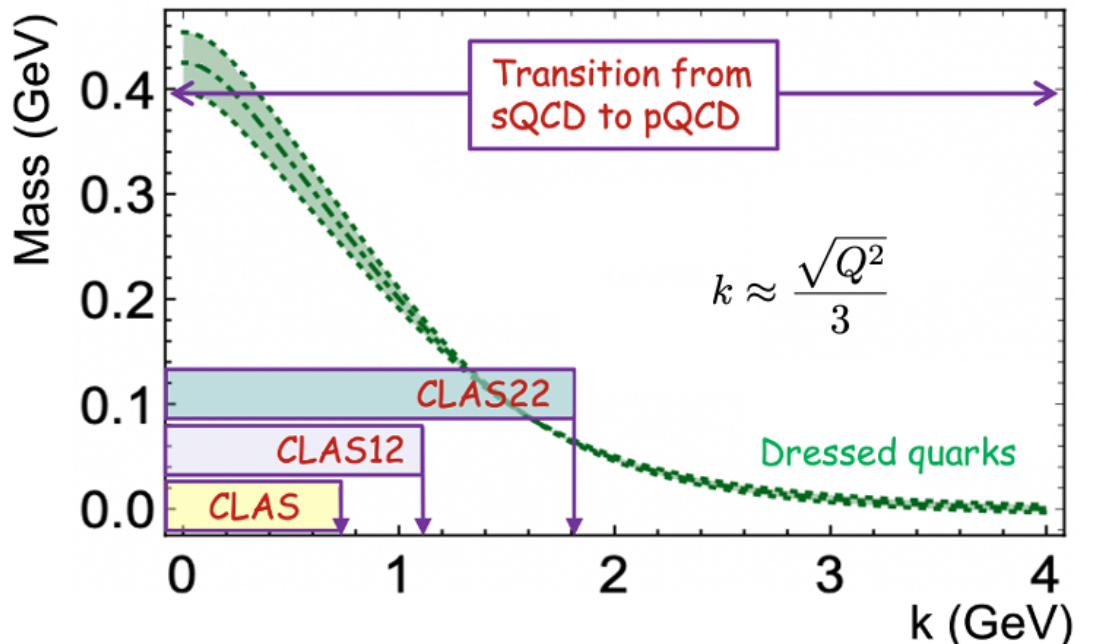}
\caption{Capabilities for mapping the momentum dependence of the dressed quark mass in studies of the $\gamma_v pN^*$ electrocouplings using data from CLAS, expected results from CLAS12, and projections for a potential CEBAF energy upgrade to 22~GeV are presented in terms of the accessible ranges of quark momenta $k$.} 
\label{kranges61222}
\end{figure}

\begin{figure}[htbp]
\centering
\includegraphics[width=0.98\columnwidth]{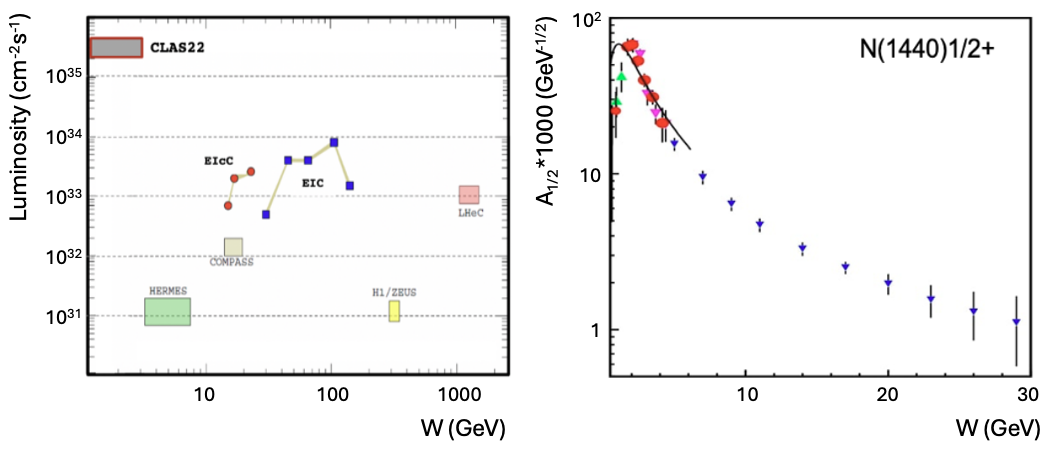}
\caption{(Left) A comparison of previous, current, and planned facilities for studying hadron structure in electroproduction, shown as a correlation plot of luminosity vs.\ $W$. (Right) Available (red and green) and projected results (blue) on the $Q^2$ evolution of the $N(1440)1/2^+$ electrocouplings~\cite{Accardi:2023chb}.}
\label{facilities}
\end{figure}

Empirical validation of the EHM paradigm as the mechanism through which QCD generates the dominant part of hadron mass requires a quantitative map of the dressed--quark mass function $M(k)$, spanning from the far infrared to the momentum domain where pQCD becomes applicable. Assuming an approximately equal sharing of the photon virtuality $Q^2$ among the three dressed quarks-—a reasonable approximation supported by the properties of baryon bound--state wavefunctions--the typical quark momentum accessible in resonance electroexcitation experiments can be estimated as $k \approx \sqrt{Q^2}/3$. The CLAS experiments of the 6--GeV era provided coverage up to $Q^2 = 5$~GeV$^2$ for most excited nucleon states with masses below 1.75~GeV~\cite{Mokeev:2022xfo, Mokeev:2024beb}. In the near future, this coverage will be extended to include resonances with masses up to 2.0~GeV for $Q^2 < 5$~GeV$^2$ \cite{mokeev-nstar24}. The corresponding range of quark momenta accessible for mapping the dressed--quark mass function from these results is illustrated in Fig.~\ref{kranges61222}.

The kinematic coverage achieved in the 6--GeV era experiments at JLab enables the exploration of the distance (momentum) domain where up to $\approx$30\% of the dressed--quark mass is generated. With the ongoing experiments in the 12--GeV era, the CLAS12 detector now provides the only facility capable of determining the electrocouplings of all prominent $N^*$ states in the previously unexplored range of photon virtualities, $5 < Q^2 < 10$~GeV$^2$. These measurements, performed across multiple exclusive meson electroproduction channels—$\pi N$, $\pi^+\pi^-p$, $\eta N$, $K\Lambda$, and $K\Sigma$—will enable a detailed mapping of the dressed--quark mass function at quark momenta corresponding to the region where roughly 50\% of the emergent hadron mass is generated.

Following a potential energy upgrade of CEBAF to 22 GeV, combined with the capability to perform exclusive meson electroproduction measurements at luminosities approaching $5 \times 10^{35}$~cm$^{-2}$s$^{-1}$, it will become possible to map the entire range of distance scales over which nearly 100\% of hadron mass and structure emerge~\cite{Accardi:2023chb}. In particular, this upgrade will enable experimental access to the $Q^2$ evolution of the $\gamma_v p N^*$ electrocouplings for the most prominent nucleon resonances, extending the coverage up to $Q^2 \approx 30$~GeV$^2$. The projected precision of the anticipated measurements for the $A_{1/2}$ electrocoupling of the Roper resonance is illustrated in Fig.~\ref{facilities} (right). These studies will provide unprecedented insight into the evolution of the dressed--quark mass across the full range of distance scales, encompassing the region where the emergent component of hadron mass is generated and culminating in the domain where the transition to pQCD is complete.

It is important to emphasize that such studies will not be feasible at any other existing or planned facility--including the future electron–ion colliders (EIC in the U.S.\ and EicC in China)--because of their luminosities being more than an order of magnitude lower than required to explore nucleon resonance electroexcitations for $Q^2 < 30$~GeV$^2$. Furthermore, their kinematic reach in the invariant mass $W$ of the final state hadronic system is largely confined to regions well above the resonance domain, as illustrated in Fig.~\ref{facilities} (left). From this perspective, a CEBAF energy upgrade to 22~GeV would establish the ultimate QCD facility at the luminosity frontier, uniquely capable of mapping the emergence of hadron mass and structure from strong QCD dynamics.

\section{Conclusion and Outlook}
\label{concl}

Nucleons and their resonances are the most fundamental three--body bound systems in Nature. A complete understanding of how QCD generates these states is therefore essential for establishing a full description of the sQCD regime. Studies of the Roper resonance structure in the 6--GeV era experiments with the CLAS detector have significantly advanced our understanding of the sQCD dynamics underlying the formation of $N^*$ states as bound systems of quarks and gluons. Analyses of the $Q^2$ evolution of the Roper electroexcitation amplitudes, extracted independently from the $\pi N$ and $\pi^+\pi^-p$ data measured with CLAS, have revealed its structure as a complex interplay between an inner core of three dressed quarks and an external meson--baryon cloud. At distance scales comparable to the hadron size ($Q^2 < 1$~GeV$^2$), the meson--baryon degrees of freedom play an important role. With increasing $Q^2$, a gradual transition occurs toward quark--core dominance, and for $Q^2 > 2$~GeV$^2$, in the case of the Roper, virtual photons penetrate the meson--baryon cloud and interact primarily with the three dressed quarks forming the resonance core.

A successful description of the Roper electrocouplings for $Q^2 > 2$~GeV$^2$ has been achieved within the CSM framework by employing the same dressed--quark mass function that reproduces the pion and nucleon elastic form factors, as well as the electroexcitation amplitudes of the $\Delta(1232)3/2^+$~\cite{Segovia:2014aza} and $\Delta(1600)3/2^+$ \cite{Mokeev:2023zhq}. This consistency strongly suggests that the dressed quark with a dynamically generated, renormalization--group--invariant, momentum--dependent mass--deduced within CSM directly from the QCD Lagrangian--serves as the active structural degree of freedom in the pion, in the ground--state nucleon, and in the excited nucleon states. Consequently, these findings provide compelling experimental support for the momentum dependence of the dressed--quark mass deduced within CSM and demonstrate the capability of mapping the dressed--quark mass function from the results on the $Q^2$--evolution of the $\gamma_v p N^*$ electrocouplings.

The kinematic reach of the CLAS detector up to $Q^2 < 5.0$~GeV$^2$ has enabled mapping of the momentum dependence of the dressed--quark mass over a limited distance range where $\approx$30\% of the hadron mass is generated. CLAS12 is currently the only facility capable of extending exclusive meson electroproduction measurements in the resonance region into the higher $Q^2$ range of 5–10~GeV$^2$. The forthcoming results on the Roper resonance electrocouplings from CLAS12 will make it possible to explore the behavior of the dressed--quark mass function in the distance domain where about 50\% of the hadron mass emerges. A future 22--GeV upgrade of CEBAF, operating with CLAS12, will enable determination of the $\gamma_v p N^*$ electrocouplings for $Q^2 < 30$~GeV$^2$~\cite{Accardi:2023chb}, providing the only foreseeable opportunity to explore the entire range of distances where hadron mass and $N^*$ structure emerge from QCD. This capability will make CEBAF@22~GeV unique and the ultimate QCD facility at the luminosity frontier.

\vskip 0.3cm

The authors would like to acknowledge the outstanding efforts of the JLab Physics Division and Accelerator staff that made it possible for the first time to gain insight into the structure of the Roper resonance from the meson electroproduction data. This work was supported in part by the the U.S. Department of Energy, Office of Science, Office of Nuclear Physics under contract DE--AC05--06OR23177.


\nocite{*}
\bibliography{References}

\begin{thebibliography}{10}

\bibitem{Roper:1964zza}
L.~David Roper.
\newblock {Evidence for a $P_{11}$ Pion-Nucleon Resonance at 556 MeV}.
\newblock {\em Phys. Rev. Lett.}, 12:340--342, 1964.

\bibitem{Roper:1965pfb}
L.~David Roper, Robert~M. Wright, and Bernard~T. Feld.
\newblock {Energy-Dependent Pion-Nucleon Phase-Shift Analysis}.
\newblock {\em Phys. Rev.}, 138:B190--B210, 1965.

\bibitem{Giannini:2015zia}
M.~M. Giannini and E.~Santopinto.
\newblock {The Hypercentral Constituent Quark Model and its Application to Baryon Properties}.
\newblock {\em Chin. J. Phys.}, 53:020301, 2015.

\bibitem{Capstick:1986ter}
Simon Capstick and Nathan Isgur.
\newblock {Baryons in a Relativized Quark Model with Chromodynamics}.
\newblock {\em Phys. Rev. D}, 34(9):2809--2835, 1986.

\bibitem{Capstick:1994ne}
Simon Capstick and B.~D. Keister.
\newblock {Baryon Current Matrix Elements in a Light Front Framework}.
\newblock {\em Phys. Rev. D}, 51:3598--3612, 1995.

\bibitem{Capstick:2000qj}
Simon Capstick and W.~Roberts.
\newblock {Quark Models of Baryon Masses and Decays}.
\newblock {\em Prog. Part. Nucl. Phys.}, 45:S241--S331, 2000.

\bibitem{Loring:2001kx}
Ulrich Loring, Bernard~C. Metsch, and Herbert~R. Petry.
\newblock {The Light Baryon Spectrum in a Relativistic Quark Model with Instanton Induced Quark Forces: The Nonstrange Baryon Spectrum and Ground States}.
\newblock {\em Eur. Phys. J. A}, 10:395--446, 2001.

\bibitem{ParticleDataGroup:2024cfk}
S.~Navas et~al.
\newblock {Review of Particle Physics}.
\newblock {\em Phys. Rev. D}, 110(3):030001, 2024.

\bibitem{Ronniger:2011td}
M.~Ronniger and B.~C. Metsch.
\newblock {Effects of a Spin-Flavour Dependent Interaction on the Baryon Mass Spectrum}.
\newblock {\em Eur. Phys. J. A}, 47:162, 2011.

\bibitem{Achenbach:2025kfx}
Patrick Achenbach, Daniel~S. Carman, Ralf~W. Gothe, Kyungseon Joo, Victor~I. Mokeev, and Craig~D. Roberts.
\newblock {Electroexcitation of Nucleon Resonances and Emergence of Hadron Mass}.
\newblock {\em Symmetry}, 17(7):1106, 2025.

\bibitem{Burkert:2017djo}
Volker~D. Burkert and Craig~D. Roberts.
\newblock {Colloquium : Roper Resonance: Toward a Solution to the Fifty Year Puzzle}.
\newblock {\em Rev. Mod. Phys.}, 91(1):011003, 2019.

\bibitem{Ding:2022ows}
Minghui Ding, Craig~D. Roberts, and Sebastian~M. Schmidt.
\newblock {Emergence of Hadron Mass and Structure}.
\newblock {\em Particles}, 6(1):57--120, 2023.

\bibitem{Arndt:1985vj}
Richard~A. Arndt, John~M. Ford, and L.~David Roper.
\newblock {Pion - Nucleon Partial Wave Analysis to 1100 MeV}.
\newblock {\em Phys. Rev. D}, 32:1085, 1985.

\bibitem{Arndt:2003if}
R.~A. Arndt, W.~J. Briscoe, I.~I. Strakovsky, R.~L. Workman, and M.~M. Pavan.
\newblock {Dispersion Relation Constrained Partial Wave Analysis of $\pi N$ Elastic and $\pi N \to \eta N$ Scattering Data: The Baryon Spectrum}.
\newblock {\em Phys. Rev. C}, 69:035213, 2004.

\bibitem{Doring:2009yv}
M.~Doring, C.~Hanhart, F.~Huang, S.~Krewald, and U.~G. Meissner.
\newblock {Analytic Properties of the Scattering Amplitude and Resonances Parameters in a Meson Exchange Model}.
\newblock {\em Nucl. Phys. A}, 829:170--209, 2009.

\bibitem{Julia-Diaz:2007qtz}
B.~Julia-Diaz, T.~S.~H. Lee, A.~Matsuyama, and T.~Sato.
\newblock {Dynamical Coupled-Channel Model of $\pi N$ Scattering in the $W \le 2$ GeV Nucleon Resonance Region}.
\newblock {\em Phys. Rev. C}, 76:065201, 2007.

\bibitem{Kamano:2010ud}
H.~Kamano, S.~X. Nakamura, T.~S.~H. Lee, and T.~Sato.
\newblock {Extraction of $P_{11}$ Resonances from $\pi N$ Data}.
\newblock {\em Phys. Rev. C}, 81:065207, 2010.

\bibitem{Segovia:2015hra}
Jorge Segovia, Bruno El-Bennich, Eduardo Rojas, Ian~C. Cloet, Craig~D. Roberts, Shu-Sheng Xu, and Hong-Shi Zong.
\newblock {Completing the Picture of the Roper Resonance}.
\newblock {\em Phys. Rev. Lett.}, 115(17):171801, 2015.

\bibitem{Liu:2016rwa}
Keh-Fei Liu.
\newblock {Baryons and Chiral Symmetry}.
\newblock {\em Int. J. Mod. Phys. E}, 26(01n02):1740016, 2017.

\bibitem{Liu:2014jua}
Keh-Fei Liu, Ying Chen, Ming Gong, Raza Sufian, Mingyang Sun, and Anyi Li.
\newblock {The Roper Puzzle}.
\newblock {\em PoS}, LATTICE2013:507, 2014.

\bibitem{Owa:2025mep}
Shiryo Owa, Derek~B. Leinweber, and Anthony~W. Thomas.
\newblock {Nucleon Resonance Structure up to 2 GeV and the Nature of the Roper Resonance}.
\newblock {\em Phys. Rev. D}, 111(11):116002, 2025.

\bibitem{Aznauryan:2009mx}
I.~G. Aznauryan et~al.
\newblock {Electroexcitation of Nucleon Resonances from CLAS Data on Single Pion Electroproduction}.
\newblock {\em Phys. Rev. C}, 80:055203, 2009.

\bibitem{Mokeev:2023zhq}
V.~I. Mokeev, P.~Achenbach, V.~D. Burkert, D.~S. Carman, R.~W. Gothe, A.~N. Hiller~Blin, E.~L. Isupov, K.~Joo, K.~Neupane, and A.~Trivedi.
\newblock {First Results on Nucleon Resonance Electroexcitation Amplitudes from $ep\to e'\pi^+\pi^-p'$ Cross Sections at $W$ = 1.4-1.7 GeV and $Q^2$ = 2.0-5.0 GeV$^2$}.
\newblock {\em Phys. Rev. C}, 108(2):025204, 2023.

\bibitem{Mokeev:2015lda}
V.~I. Mokeev, V.~D. Burkert, D.~S. Carman, L.~Elouadrhiri, G.~V. Fedotov, E.~N. Golovatch, R.~W. Gothe, K.~Hicks, B.~S. Ishkhanov, E.~L. Isupov, and Iu. Skorodumina.
\newblock {New Results from the Studies of the $N(1440)1/2^+$, $N(1520)3/2^-$, and $\Delta(1620)1/2^-$ Resonances in Exclusive $ep \to e'p'\pi^+ \pi^-$ Electroproduction with the CLAS Detector}.
\newblock {\em Phys. Rev. C}, 93(2):025206, 2016.

\bibitem{Mokeev:2012vsa}
V.~I. Mokeev et~al.
\newblock {Experimental Study of the $P_{11}(1440)$ and $D_{13}(1520)$ Resonances from CLAS Data on $ep \to e' \pi^+ \pi^- p'$}.
\newblock {\em Phys. Rev. C}, 86:035203, 2012.

\bibitem{Mokeev:2022xfo}
Victor~I. Mokeev and Daniel~S. Carman.
\newblock {Photo- and Electrocouplings of Nucleon Resonances}.
\newblock {\em Few Body Syst.}, 63(3):59, 2022.

\bibitem{Chesnokov:2022gjb}
V.~V. Chesnokov, A.~A. Golubenko, B.~S. Ishkhanov, and V.~I. Mokeev.
\newblock {CLAS Database for Studies of the Structure of Hadrons in Electromagnetic Processes}.
\newblock {\em Phys. Part. Nucl.}, 53(2):184--190, 2022.

\bibitem{CLAS:DB}
{CLAS Collaboration}.
\newblock {CLAS Physics Database}.
\newblock \url{https://clasweb.jlab.org/physicsdb/}.

\bibitem{Aznauryan:2011qj}
I.~G. Aznauryan and V.~D. Burkert.
\newblock {Electroexcitation of Nucleon Resonances}.
\newblock {\em Prog. Part. Nucl. Phys.}, 67:1--54, 2012.

\bibitem{gvnstarp:gwu}
{GWU}.
\newblock {SAID Database}.
\newblock \url{https://jbw.phys.gwu.edu/}.

\bibitem{gvnstarp:dat}
{V.I. Mokeev}.
\newblock {Nucleon Resonance Photo-/Electrocouplings Determined from Analyses of Experimental Data on Exclusive Meson Electroproduction off Protons}.
\newblock \url{https://userweb.jlab.org/~mokeev/resonance\_electrocouplings23/}.

\bibitem{HillerBlin:2019jgp}
A.~N. Hiller~Blin et~al.
\newblock {Nucleon Resonance Contributions to Unpolarized Inclusive Electron Scattering}.
\newblock {\em Phys. Rev. C}, 100(3):035201, 2019.

\bibitem{Wang:2024byt}
Yu-Fei Wang, Michael D{\"o}ring, Jackson Hergenrather, Maxim Mai, Terry Mart, Ulf-G. Mei{\ss}ner, Deborah R{\"o}nchen, and Ronald Workman.
\newblock {Global Data-Driven Determination of Baryon Transition Form Factors}.
\newblock {\em Phys. Rev. Lett.}, 133(10):101901, 2024.

\bibitem{Kamano:2016bko}
H.~Kamano, S.~X. Nakamura, T.~S.~H. Lee, and T.~Sato.
\newblock {Isospin Decomposition of the $\gamma^* N \to N^*$ Transitions as Input for Constructing Models of Neutrino-Induced Reactions in the Nucleon Resonance Region}.
\newblock 10 2016.

\bibitem{CLAS:2009tyz}
M.~Dugger et~al.
\newblock {$\pi^+$ Photoproduction on the Proton for Photon Energies from 0.725 to 2.875 GeV}.
\newblock {\em Phys. Rev. C}, 79:065206, 2009.

\bibitem{Aznauryan:2018okk}
I.~G. Aznauryan and V.~D. Burkert.
\newblock {Electroexcitation of Nucleon Resonances in a Light-Front Relativistic Quark Model}.
\newblock {\em Few Body Syst.}, 59(5):98, 2018.

\bibitem{Obukhovsky:2011sc}
I.~T. Obukhovsky, Amand Faessler, D.~K. Fedorov, Thomas Gutsche, and Valery~E. Lyubovitskij.
\newblock {Electroproduction of the Roper Resonance on the Proton: The Role of the Three-Quark Core and the Molecular $N\sigma$ Component}.
\newblock {\em Phys. Rev. D}, 84:014004, 2011.

\bibitem{Barabanov:2020jvn}
M.~Yu. Barabanov et~al.
\newblock {Diquark Correlations in Hadron Physics: Origin, Impact and Evidence}.
\newblock {\em Prog. Part. Nucl. Phys.}, 116:103835, 2021.

\bibitem{Segovia:2014aza}
Jorge Segovia, Ian~C. Cloet, Craig~D. Roberts, and Sebastian~M. Schmidt.
\newblock {Nucleon and $\Delta$ Elastic and Transition Form Factors}.
\newblock {\em Few Body Syst.}, 55:1185--1222, 2014.

\bibitem{Aznauryan:2012ec}
I.~G. Aznauryan and V.~D. Burkert.
\newblock {Nucleon Electromagnetic Form Factors and Electroexcitation of Low Lying Nucleon Resonances in a Light-Front Relativistic Quark Model}.
\newblock {\em Phys. Rev. C}, 85:055202, 2012.

\bibitem{Bauer:2014cqa}
T.~Bauer, S.~Scherer, and L.~Tiator.
\newblock {Electromagnetic Transition Form Factors of the Roper Resonance in Effective Field Theory}.
\newblock {\em Phys. Rev. C}, 90(1):015201, 2014.

\bibitem{Krehl:1999km}
O.~Krehl, C.~Hanhart, S.~Krewald, and J.~Speth.
\newblock {What is the Structure of the Roper Resonance?}
\newblock {\em Phys. Rev. C}, 62:025207, 2000.

\bibitem{Speth:2000zf}
J.~Speth, O.~Krehl, S.~Krewald, and C.~Hanhart.
\newblock {The Structure of the Roper Resonance}.
\newblock {\em Nucl. Phys. A}, 680:328--334, 2000.

\bibitem{Mokeev:2008iw}
Viktor~I. Mokeev, Volker~D. Burkert, Tsung-Shung~H. Lee, Latifa Elouadrhiri, Gleb~V. Fedotov, and Boris~S. Ishkhanov.
\newblock {Model Analysis of the $p \pi^+ \pi^-$ Electroproduction Reaction on the Proton}.
\newblock {\em Phys. Rev. C}, 80:045212, 2009.

\bibitem{Ripani:2000rz}
M.~Ripani et~al.
\newblock {A Phenomenological Description of $\pi^- \Delta^{++}$ Photoproduction and Electroproduction in Nucleon Resonance Region: Erratum}.
\newblock 10 2000.

\bibitem{Wang:2023snv}
Yu-Fei Wang, Ulf-G. Mei{\ss}ner, Deborah R{\"o}nchen, and Chao-Wei Shen.
\newblock {Examination of the Nature of the $N^*$ and $\Delta$ Resonances via Coupled-Channels Dynamics}.
\newblock {\em Phys. Rev. C}, 109(1):015202, 2024.

\bibitem{Zou:2003zn}
B.~S. Zou and D.~O. Riska.
\newblock {The Roper Resonance as a Hybrid of Three-Quark and Five-Quark Configurations}.
\newblock {\em Phys. Rev. D}, 69:034004, 2004.

\bibitem{Englert:2014zpa}
Fran{\c{c}}ois Englert.
\newblock {Nobel Lecture: The BEH Mechanism and its Scalar Boson}.
\newblock {\em Rev. Mod. Phys.}, 86(3):843, 2014.

\bibitem{Higgs:2014aqa}
Peter~W. Higgs.
\newblock {Nobel Lecture: Evading the Goldstone Theorem}.
\newblock {\em Rev. Mod. Phys.}, 86(3):851, 2014.

\bibitem{ATLAS:2012yve}
Georges Aad et~al.
\newblock {Observation of a New Particle in the Search for the Standard Model Higgs Boson with the ATLAS Detector at the LHC}.
\newblock {\em Phys. Lett. B}, 716:1--29, 2012.

\bibitem{CMS:2012qbp}
{CMS Collaboration}.
\newblock {Observation of a New Boson at a Mass of 125 GeV with the CMS Experiment at the LHC}.
\newblock {\em Phys. Lett. B}, 716:30--61, 2012.

\bibitem{Binosi:2025kpz}
Daniele Binosi, Craig~D. Roberts, and Zhao-Qian Yao.
\newblock {Hadron Structure: Perspective and Insights}.
\newblock {\em PoS}, QCHSC24:001, 2025.

\bibitem{Horn:2016rip}
Tanja Horn and Craig~D. Roberts.
\newblock {The Pion: An Enigma within the Standard Model}.
\newblock {\em J. Phys. G}, 43(7):073001, 2016.

\bibitem{Roberts:2020hiw}
Craig~D Roberts.
\newblock {Empirical Consequences of Emergent Mass}.
\newblock {\em Symmetry}, 12(9):1468, 2020.

\bibitem{Roberts:2021xnz}
Craig~D. Roberts.
\newblock {On Mass and Matter}.
\newblock {\em AAPPS Bull.}, 31:6, 2021.

\bibitem{Roberts:2021nhw}
Craig~D. Roberts, David~G. Richards, Tanja Horn, and Lei Chang.
\newblock {Insights into the Emergence of Mass from Studies of Pion and Kaon Structure}.
\newblock {\em Prog. Part. Nucl. Phys.}, 120:103883, 2021.

\bibitem{Cui:2019dwv}
Zhu-Fang Cui, Jin-Li Zhang, Daniele Binosi, Feliciano de~Soto, C{\'e}dric Mezrag, Joannis Papavassiliou, Craig~D. Roberts, Jose Rodr{\'\i}guez-Quintero, Jorge Segovia, and Savvas Zafeiropoulos.
\newblock {Effective Charge from Lattice QCD}.
\newblock {\em Chin. Phys. C}, 44(8):083102, 2020.

\bibitem{Deur:2022msf}
A.~Deur, V.~Burkert, J.~P. Chen, and W.~Korsch.
\newblock {Experimental Determination of the QCD Effective Charge $\alpha_{g_1}(Q)$}.
\newblock {\em Particles}, 5:171, 2022.

\bibitem{Deur:2023dzc}
Alexandre Deur, Stanley~J. Brodsky, and Craig~D. Roberts.
\newblock {QCD Running Couplings and Effective Charges}.
\newblock {\em Prog. Part. Nucl. Phys.}, 134:104081, 2024.

\bibitem{Schwinger:1962tn}
Julian~S. Schwinger.
\newblock {Gauge Invariance and Mass}.
\newblock {\em Phys. Rev.}, 125:397--398, 1962.

\bibitem{Schwinger:1962tp}
Julian~S. Schwinger.
\newblock {Gauge Invariance and Mass. 2.}
\newblock {\em Phys. Rev.}, 128:2425--2429, 1962.

\bibitem{Cloet:2013jya}
Ian~C. Cloet and Craig~D. Roberts.
\newblock {Explanation and Prediction of Observables using Continuum Strong QCD}.
\newblock {\em Prog. Part. Nucl. Phys.}, 77:1--69, 2014.

\bibitem{Wilson:2011aa}
D.~J. Wilson, I.~C. Cloet, L.~Chang, and C.~D. Roberts.
\newblock {Nucleon and Roper Electromagnetic Elastic and Transition Form Factors}.
\newblock {\em Phys. Rev. C}, 85:025205, 2012.

\bibitem{Segovia:2013uga}
Jorge Segovia, Chen Chen, Ian~C. Clo{\"e}t, Craig~D. Roberts, Sebastian~M. Schmidt, and Shaolong Wan.
\newblock {Elastic and Transition Form Factors of the $\Delta(1232)$}.
\newblock {\em Few Body Syst.}, 55:1--33, 2014.

\bibitem{CLAS:2006fml}
I.~G. Aznauryan, V.~D. Burkert, H.~Egiyan, K.~Joo, R.~Minehart, L.~C. Smith, and {CLAS Collaboration}.
\newblock {Electroexcitation of Nucleon Resonances from CLAS Data on Single Pion Electroproduction}.
\newblock {\em Phys. Rev. C}, 73:025204, 2006.

\bibitem{Villano:2009sn}
A.N. Villano et~al.
\newblock {Neutral Pion Electroproduction in the Resonance Region at High $Q^2$}.
\newblock {\em Phys. Rev. C}, 80:035203, 2009.

\bibitem{CLAS:2012wxw}
V.~I. Mokeev et~al.
\newblock {Experimental Study of the $P_{11}(1440)$ and $D_{13}(1520)$ Resonances from CLAS data on $ep \to e'\pi^+ \pi^- p'$}.
\newblock {\em Phys. Rev. C}, 86:035203, 2012.

\bibitem{NA7:1986vav}
S.~R. Amendolia et~al.
\newblock {A Measurement of the Space - Like Pion Electromagnetic Form-Factor}.
\newblock {\em Nucl. Phys. B}, 277:168, 1986.

\bibitem{Horn:2007ug}
T.~Horn et~al.
\newblock {Scaling Study of the Pion Electroproduction Cross Sections and the Pion Form Factor}.
\newblock {\em Phys. Rev. C}, 78:058201, 2008.

\bibitem{JeffersonLab:2008jve}
G.~M. Huber et~al.
\newblock {Charged Pion Form-Factor Between $Q^2 = 0.60$ GeV$^2$ and 2.45 GeV$^2$. II. Determination of, and Results for, the Pion Form-Factor}.
\newblock {\em Phys. Rev. C}, 78:045203, 2008.

\bibitem{JeffersonLabHallA:2001qqe}
O.~Gayou et~al.
\newblock {Measurement of $G_{E_p}/G_{M_p}$ in $\vec{e} p \to e \vec{p}$ to $Q^2$ = 5.6 GeV$^2$}.
\newblock {\em Phys. Rev. Lett.}, 88:092301, 2002.

\bibitem{JeffersonLabHallA:1999epl}
M.~K. Jones et~al.
\newblock {$G_{E_p}/G_{M_p}$ Ratio by Polarization Transfer in $\vec ep \to e\vec p$}.
\newblock {\em Phys. Rev. Lett.}, 84:1398--1402, 2000.

\bibitem{Punjabi:2005wq}
V.~Punjabi et~al.
\newblock {Proton Elastic Form-Factor Ratios to $Q^2$ = 3.5 GeV$^2$ by Polarization Transfer}.
\newblock {\em Phys. Rev. C}, 71:055202, 2005.
\newblock [Erratum: Phys.Rev.C 71, 069902 (2005)].

\bibitem{Cui:2020rmu}
Zhu-Fang Cui, Chen Chen, Daniele Binosi, Feliciano de~Soto, Craig~D Roberts, Jose Rodr{\'\i}guez-Quintero, Sebastian~M Schmidt, and Jorge Segovia.
\newblock {Nucleon Elastic Form Factors at Accessible Large Spacelike Momenta}.
\newblock {\em Phys. Rev. D}, 102(1):014043, 2020.

\bibitem{Bender:1996bb}
A.~Bender, C.~D. Roberts, and L.~Von Smekal.
\newblock {Goldstone Theorem and Diquark Confinement Beyond Rainbow-Ladder Approximation}.
\newblock {\em Phys. Lett. B}, 380:7--12, 1996.

\bibitem{Munczek:1994zz}
H.~J. Munczek.
\newblock {Dynamical Chiral Symmetry Breaking, Goldstone's Theorem and the Consistency of the Schwinger-Dyson and Bethe-Salpeter Equations}.
\newblock {\em Phys. Rev. D}, 52:4736--4740, 1995.

\bibitem{Aznauryan:2016wwm}
I.~G. Aznauryan and V.~D. Burkert.
\newblock {Nucleon Electromagnetic Form Factors and Electroexcitation of Low-Lying Nucleon Resonances in a Light-Front Relativistic Quark Model}.
\newblock {\em Phys. Rev. C}, 92:035211, 2015.

\bibitem{Yao:2024uej}
Zhao-Qian Yao, Daniele Binosi, Zhu-Fang Cu, and Craig~D. Roberts.
\newblock {Nucleon Charge and Magnetisation Distributions: Flavour Separation and Zeroes}.
\newblock 3 2024.

\bibitem{Yao:2024drm}
Zhao-Qian Yao, Daniele Binosi, and Craig~D. Roberts.
\newblock {Onset of Scaling Violation in Pion and Kaon Elastic Electromagnetic Form Factors}.
\newblock {\em Phys. Lett. B}, 855:138823, 2024.

\bibitem{Accardi:2023chb}
A.~Accardi et~al.
\newblock {Strong Interaction Physics at the Luminosity Frontier with 22 GeV Electrons at Jefferson Lab}.
\newblock {\em Eur. Phys. J. A}, 60:173, 2024.

\bibitem{Mokeev:2024beb}
V.~I. Mokeev and D.~S. Carman.
\newblock {Insight into Emergence of Hadron Mass from $N^*$ Electroexcitation Amplitudes}.
\newblock {\em Nuovo Cim. C}, 47:216, 2024.

\bibitem{mokeev-nstar24}
V.~I. Mokeev.
\newblock {$N^*$ Structure and the Emergence of Hadron Mass}.
\newblock In {\em Proceedings of the NSTAR2024 --- 14th International Workshop on the Physics of Excited Nucleons}, York, UK, June 2024.
\newblock 17--21 June 2024.

\end{thebibliography}
\end{document}